\documentclass[sigconf,nonacm]{acmart}

\pdfoutput=1

\AtBeginDocument{%
  \providecommand\BibTeX{{%
    \normalfont B\kern-0.5em{\scshape i\kern-0.25em b}\kern-0.8em\TeX}}}

\usepackage{enumitem}

\setcopyright{none}
\copyrightyear{2019}

\begin{document}

\title{In AI We Trust? Factors That Influence Trustworthiness of AI-infused Decision-Making Processes}

\author{Maryam Ashoori}
\authornote{Both authors contributed equally to this research.}
\affiliation{%
  \institution{IBM Research AI}
  \city{Yorktown Heights}
  \state{NY}
  \postcode{10598}
}
\email{maryam@us.ibm.com}

\author{Justin D. Weisz}
\authornotemark[1]
\affiliation{%
  \institution{IBM Research AI}
  \city{Yorktown Heights}
  \state{NY}
  \postcode{10598}
}
\email{jweisz@us.ibm.com}


\begin{abstract}
Many decision-making processes have begun to incorporate an AI element, including prison sentence recommendations~\cite{angwin2016machine, hao2019jail}, college admissions~\cite{pangburn2019schools}, hiring~\cite{dattner2019hiring}, and mortgage approval~\cite{murawski2019mortgage}. In all of these cases, AI models are being trained to help human decision makers reach accurate and fair judgments, but little is known about what factors influence the extent to which people consider an AI-infused decision-making process to be trustworthy. We aim to understand how different factors about a decision-making process, and an AI model that supports that process, influences peoples' perceptions of the trustworthiness of that process. We report on our evaluation of how seven different factors -- decision stakes, decision authority, model trainer, model interpretability, social transparency, and model confidence -- influence ratings of trust in a scenario-based study.
\end{abstract}



\maketitle

\section{Introduction}
Recent technological breakthroughs in the areas of artificial intelligence and machine learning have sparked a surge of interest in the practice of data science. Fueled by data, data scientists are training AI-based models that help provide recommendations to human decision makers in a variety of domains. These AI-infused decision-making processes include high-stakes domains such as prison sentence recommendation~\cite{hao2019jail}, credit decisions~\cite{murawski2019mortgage}, and hiring~\cite{dattner2019hiring}, as well as more familiar lower-stakes domains such as personalized shopping~\cite{linden2003amazon} and music recommendation~\cite{koenigstein2011yahoo}.

Despite the growing popularity of training AI-based models to aid human decision makers, little is known about peoples' perceptions of trust in an AI-infused decision-making process. What characteristics of such a process makes it trustworthy? What kinds of information are needed about the process to make people trust that it will produce a reasonable outcome?

There are many questions that can be asked about how an AI model is incorporated into a decision-making process. Who owns responsibility for the final decision, AI or human? How much does one know about the data that went into the AI model? Is the model interpretable, making it easy to understand why it made a particular recommendation, or is it a ``black box'' that gives no information about how it came to its conclusion? (e.g.,~\cite{rudin2018please}). How does trust depend on the stakes of the decision, whether it is lower stakes such as choosing a song to listen to or a restaurant to try, or higher stakes such as whether one obtains a mortgage loan or how long one spends in prison after having been convicted of a crime?

We report on our work in understanding the impact of factors such as these on the trustworthiness of an AI-infused decision-making process. We begin by giving an overview of prior work on trust and what it means to trust an AI system. We then describe our crowdsourced, scenario-based study to elicit peoples' ratings of their trust if different AI-infused decision-making processes. Our paper makes the following contributions to the HCI \& AI communities:

\begin{itemize}[leftmargin=*]
    \item We perform a comprehensive quantitative analysis of the effect of seven different factors and their impact on four different facets of trust, showing strong statistical evidence that the use of interpretable models and the inclusion of information about how AI models were trained and tested leads to increased trust ratings,
    \item We provide a nuanced depiction of the myriad of individual differences of opinion people have on the trustworthiness of different kinds of AI-infused decision making processes,
    \item We challenge the community to identify additional factors that impact feelings of trust toward AI systems, as well as think critically about whether trust in AI systems is always a desirable characteristic.
\end{itemize}

We expect our results to influence the development of trusted AI systems by highlighting the importance of soliciting input not only from the data scientists and engineers who build these systems, but also by the people who will use these systems and those affected by decisions made by these systems.

\section{What is Trustworthy AI?}
We describe prior work in the area of trust, with an emphasis on the multitude of facets that comprise trust and how those facets have been operationalized and measured.

\subsection{Definition of Trust}
Trust is a multi-dimensional concept that has been extensively examined in a wide range of fields. For example, it is widely acknowledged that trust is a significant component in cooperative relationships, expressed as ``an expectancy held by an individual or a group that the word, promise... of another individual or group can be relied upon.''~\cite{rotter1967new}. Trust has also been examined in the context of automated systems, and Adams et al.~\cite{adams2003trust} provides a summary of how trust in automated systems compares to, and differs from, interpersonal trust. But, despite our intuitive notions of what ``trust'' is, the definitions and conceptualizations of trust widely vary. One of the major inconsistencies between trust definitions is whether trust is considered to be an attitude~\cite{rotter1967new}, an intention~\cite{mayer1995integrative}, or a behavior~\cite{adams2003trust}.

In this paper, we adopt a reconciliation of the conflicting definitions of trust proposed by Lee and See~\cite{lee2004trust}. They define trust as an attitude, rather than as a belief, intention, or behavior: ``[trust is] the attitude that an agent will help achieve an individual's goals in a situation characterized by uncertainty and vulnerability.'' Since its publication, this definition of trust has been the most widely used in empirical studies of trust in human-machine interactions~\cite{french2018trust}. In the context of AI-infused decision making, Madsen and Gregor~\cite{madsen2000measuring} have defined trust as ``the extent to which a user is confident in, and willing to act on the basis of, the recommendations, actions, and decisions of an artificially intelligent decision aid.'' By adopting the attitudinal viewpoint, we are able to evaluate trust in the context of peoples' attitudes toward hypothetical situations, rather than having to base our evaluations of trust on specific behaviors.

\subsection{Facets of Trust}
One of the greatest challenges facing research in the area of trustworthy AI is to precisely define and measure trust and its role in mediating relationships between people and machines. Trust is highly dependent on circumstances~\cite{balfe2018understanding}, and previous measurements of trust (e.g.,~\cite{cahour2009does, dzindolet2003role, jian2000foundations, merritt2011affective}) have largely relied on subjective assessments of what they believed influenced trust in specific situations. Much of the prior work on measuring trust in automated systems has focused on evaluating the trustworthiness of the decision rather than evaluating the trustworthiness of the decision-making process itself. In our study, we focus on the latter by leveraging established measures of different facets of trust to evaluate the trustworthiness of an AI-infused decision-making \emph{process}.

Adams et al.~\cite{adams2003trust} analyzed trust within the context of several different factors: usefulness, reliability, accuracy, understandability, joy of use, and ease of use. Cahour \& Forzy~\cite{cahour2009does} evaluated peoples' confidence in AI decisions based on predictability, reliability, safety, and efficiency. Muir~\cite{muir1994trust}, Madsen \& Gregor~\cite{madsen2000measuring}, and Kelly et al.~\cite{kelly2003guidelines} argued that trust formation also depended upon machine competence (i.e., the extent to which it does its job properly).

One of the most widely used trust scales in the field of human factors is that of Jian et al.~\cite{jian2000foundations}, which analyzes six facets of trust: fidelity, loyalty, reliability, security, integrity, and familiarity. However, as stated by Hoffman et al.~\cite{hoffman2018metrics}, some of these facets are problematic when applied to a machine. For example, the concept that a machine can act with ``integrity'' is not explicated. Focusing on intelligent decision aids, Madsen \& Gregor~\cite{madsen2000measuring} analyzed trust within five facets: reliability, technical competence, understandability, faith, and personal attachment. Balfe et al.~\cite{balfe2018understanding} expanded upon this set by examining additional dimensions of reliability, robustness, understandability, competence, feedback, dependability, personal attachment, predictability, and faith. Hoffman et al.~\cite{hoffman2018metrics} analyzed a similar set of dimensions for an explainable AI system by asking users whether they were confident in the system, and whether the system was predictable, reliable, efficient, and believable.

From this wealth of prior research in trust of automated systems, we conclude that ``trust'' must be examined in a multi-dimensional way. Based on our goal of evaluating trust in an AI-infused decision-making process, we find these facets of trust to be most relevant: overall trustworthiness, reliability, technical competence, understandability, and personal attachment.

\section{Trust Boundaries in AI-infused Decision Making}
Our primary research question is to identify how different aspects of an AI-infused decision-making process influence peoples' feelings of trust in that process. We seek to find the boundary between processes that are considered trustworthy vs. not trustworthy, similar to Dodge et al's study of how explanations impact peoples' judgments of the fairness of an AI algorithm~\cite{dodge2019explaining}. We examine the following factors to determine the shape of this boundary:

\begin{itemize}[leftmargin=*]
    \item \textbf{Stakes}. What are the consequences of this decision? We consider both lower-stakes decisions (e.g. meal planning) and higher-stakes decisions (e.g. prison sentencing), especially in light of the increasing use of AI models for higher-stakes decisions (e.g.,~\cite{rudin2018please, hao2019jail, murawski2019mortgage}).
    \item \textbf{Decider}. Who is responsible for making the ultimate decision? We compare decisions made entirely by AI to decisions made by a human supported by AI guidance.
    \item \textbf{Trainer}. How was the AI trained? Given the current popularity of automated training methods for AI (e.g.,~\cite{feurer2015efficient, zoller2019survey}), we consider AI models trained by human data scientists vs. those trained by automated AI.
    \item \textbf{Model Interpretability}. Some kinds of AI models are interpretable, such as decision trees or rule-based scoring systems~\cite{bertsimas2019price,rudin2018please}. For interpretable models, the process by which the model arrived at a recommendation can be examined and understood. Other models are considered to be ``black boxes,'' such as deep neural networks, whose inner workings do not give insight into how a recommendation was made. Work by Glass et al.~\cite{glass2008toward} suggests that availability of explanations is an important aspect for establishing trust of an intelligent agent. Although we recognize that explanations can be generated for black box models (e.g.,~\cite{ribeiro2016should, arya2018aix}), we consider interpretability to be a more desirable characteristic of a model given recent criticisms of black box models~\cite{rudin2018please}.
    \item \textbf{Train \& Test Set Description}. Does the model come with a description of how it was trained and tested? Researchers and other advocates for fairness in AI have called for transparency in how AI models have been trained in order to enable others to evaluate their trustworthiness~\cite{hind2018increasing, bellamy2019aif}. However, we recognize that some models are kept secret over concerns of being proprietary or to maintain a competitive advantage~\cite{rudin2018age}.
    \item \textbf{Social Transparency}. Are decisions made for others transparent? Seeing how a model makes decisions for other people may be an indicator of whether that model exhibits bias.
    \item \textbf{Model Confidence}. Is the confidence a model has in its recommendation visible? Many types of AI models provide the ability to give an estimate of their confidence when making a recommendation.
\end{itemize}

\section{Evaluating Trust in AI-Infused Decision Making}

\subsection{Methodology}
To evaluate how each of the above factors affects trust of an AI-infused decision-making process, we used a scenario-based approach in which a decision-making process is described, and participants rate their attitudes toward it. Similar scenario-based designs have been used to evaluate peoples' perceptions of AI (e.g.,~\cite{dodge2019explaining}) and IoT (e.g.,~\cite{uchidiuno2018privacy}) systems.

For each unique combination of factors, we constructed a scenario from a bank of pre-written sentences that described each level of each factor. There were 128 unique combinations: 2 (stakes) x 2 (decider) x 2 (trainer) x 2 (interpretability) x 2 (train/test description) x 2 (transparency) x 2 (confidence). Table~\ref{tab:examples} shows two example scenarios showing descriptions of each level of each factor. For clarity, we refer to these factors as the scenario factors.

\subsection{Participants}
In order to sample a wider variety of viewpoints than those available in our institution, we employed crowdsourcing methods. Crowdsourcing, while popular for collecting ground truth labels used to train AI systems, has also been used as a means of sampling opinions on new technologies~\cite{uchidiuno2018privacy} or of gaining preliminary feedback on demonstrations of new technology~\cite{weisz2019bigbluebot}.

We recruited 362 participants from Mechanical Turk, with the requirements of being 18 years or older and speaking English fluently. As our scenarios are written in English, we did not want a lack of understanding to confound participants' ratings.

\subsection{Procedure}
Participants reviewed a description of the study before deciding whether to participate. Each participant was then presented with two scenarios -- one lower stakes and one higher stakes -- in random order, with all other scenario factors determined at random.

We paid participants \$1.25 USD for our task, informed by pretesting that suggested the task took approximately 8-10 minutes to complete, and our desire to pay according to U.S. minimum wage (\$7.25 USD/hr). Participants actually spent between 3 and 60 minutes completing the task, with a median of approximately 16 minutes. About 80\% of participants completed the task in less than 30 minutes. Thus, our actual payout for the median participant corresponded to \$4.69 USD per hour.

\begin{table}[htp]
    \small
    \centering
    \begin{tabular}{p{8cm}}
        \emph{a. Stakes: high, Decider: AI, Trainer: automated AI, Interpretability: interpretable, Train \& Test Set Description: present, Transparency: present, Confidence: present} \newline  
        \\
        A person is convicted of a crime and the duration of the prison sentence needs to be determined. Here's what you know about how the prison sentence will be determined:
        \begin{itemize}[leftmargin=*]
            \item The duration of the prison sentence will be completely determined by an AI model.
            \item The AI model was built automatically, without human intervention, from historical data on crimes and prison sentences.
            \item By examining the model, one is able to understand exactly why a decision was made and what factors were used in making that decision.
            \item Information about the data used to train the model, including where that data came from, how much data was used for training, how much data was used to evaluate the accuracy of the model, and that model's accuracy will be made available.
            \item Information about other prison sentence recommendations made for other defendants will be made available.
            \item Information about how confident the AI model is in its prison sentence recommendation will be available.
        \end{itemize} 
        \\
        \emph{b. Stakes: low, Decider: human, Trainer: human, Interpretability: blackbox, Train \& Test Set Description: absent, Transparency: absent, Confidence: absent} \newline  
        \\
        A meal service customer needs to have a meal plan created. Here's what you know about how the meal plan will be created:
        \begin{itemize}[leftmargin=*]
            \item A chef will use an AI model to help determine the meal plan, but the final decision will be made by the chef.
            \item The AI model was created by a team of data scientists, who trained it using historical data on meal preferences and nutritional information.
            \item By examining the model, one is not able to understand exactly why a decision was made nor what factors were used in making that decision.
            \item No information will be made available about where the data
            used to train the model came from, how much data was used for training, how much data was used to evaluate the accuracy of the model, nor that model's accuracy.
            \item Information about other meal plan recommendations made for other customers will not be made available.
            \item Information about how confident the AI model is in its meal plan recommendation will not be available.
        \end{itemize}
    \end{tabular}
    \caption{Examples of higher- and lower-stakes scenarios.}
    \label{tab:examples}
\end{table}

\subsection{Measures}
As discussed earlier, trust is a multi-dimensional concept. Thus, our evaluation of trust focused on several dimensions we felt were most relevant:

\begin{itemize}[leftmargin=*]
    \itemsep0em
    \item \textbf{Overall trustworthiness}: the process ought to be trusted,
    \item \textbf{Reliability}: the process results in consistent outcomes,
    \item \textbf{Technical competence}: AI is used appropriately and correctly,
    \item \textbf{Understandability}: participants understood how the process works, and
    \item \textbf{Personal attachment}: participants liked the process. 
\end{itemize}

As we did not find a prior scale to evaluate overall trustworthiness, we developed our own 4-item scale. We adapted the existing scales of Madsen \& Gregor~\cite{madsen2000measuring} for measuring reliability, technical competence, and understandability, and Hoffman et al.~\cite{hoffman2018metrics} for personal attachment. Our adaptations included rewording and/or dropping items to make sense in the context of our scenarios. All items were rated on 4-point Likert scales: ``Strongly disagree,'' ``Disagree,'' ``Agree,'' and ``Strongly agree.'' Details of these scales are shown in Table~\ref{tab:scales}.

\begin{table}[ht]
    \small
    \centering
    \begin{tabular}{p{8cm}} 
        \textbf{Trustworthiness} ($\alpha = 0.82$) \newline
        1. This decision-making process is trustworthy \newline
        2. I would change one or more aspects of this decision-making process to make it trustworthy* \newline
        3. This decision-making process will produce a fair outcome for the person affected by the decision \newline
        4. The decision maker needs more information about how the AI model was trained and tested in order to trust the process* \newline
        \\
        \textbf{Reliability ($\alpha = 0.74$)} \newline
        1. This decision-making process would always make the same recommendation under the same conditions \newline
        2. The outcome of this decision will be consistent with other decisions made for other people \newline
        \\
        \textbf{Technical Competence} ($\alpha = 0.86$) \newline
        1. The use of an AI model is appropriate in this scenario \newline
        2. This decision will be made based on reliable information \newline
        3. I trust that the technical implementation of the AI model is correct \newline
        \\
        \textbf{Understandability**} ($\alpha = 0.11$) \newline
        1. It is easy to understand what this decision-making process does \newline
        2. I understand how this decision-making process works \newline
        \\
        \textbf{Personal Attachment} ($\alpha = 0.90$) \newline
        1. I am confident in this decision-making process. I feel that it works well \newline
        2. I am wary of this decision-making process* \newline
        3. I like this decision-making process   \newline      
        \\
    \end{tabular}
    \caption{Scales used to evaluate different facets of trust. Reliabilities are reported as Cronbach's $\alpha$. (*) Items with an asterisk were reverse-coded. (**) Due to poor reliability, we exclude this facet from our analysis.}
    \label{tab:scales}
\end{table}

In addition to Likert scale items, participants were asked to qualitatively describe their feelings toward the given decision-making process: why or why isn't it trustworthy, what additional information is needed to make it trustworthy, and how would they change the process to be more trustworthy?

We also included a manipulation check in the form of attention questions to ensure the quality of survey responses on Mechanical Turk, as recommended by Kittur et al.~\cite{kittur2008crowdsourcing}. Each participant was asked two questions about the specific details of each scenario (4 questions total) to ensure they read and understood it. When receiving results from Mechanical Turk, we noticed that about 69\% of participants answered all four attention questions correctly, and 88\% of participants answered three or more questions correctly. Thus, to avoid discarding a large proportion of work, we excluded participants who answered fewer than three attention questions correctly; scenarios that were excluded were re-submitted until the attention questions met our threshold. Ultimately, we discarded work from 42 participants (11.6\%), leaving us with a final data set of 640 scenario ratings (5 per scenario) from N=320 participants.

Finally, we included a question to gauge participants' perception of the stakes of each scenario to understand whether our intent of lower vs. higher stakes matched their interpretation: ``If this decision-making process results in an unfavorable outcome, how stressful will it be for the affected person?'' This question was measured on a 7-point scale (``Not very stressful'' to ``Significantly stressful'').

\section{Analysis of Scenario Factors on Trust Facets}
We performed several kinds of analysis to determine where the boundaries of trust lay amongst the different factors. First, we conducted confirmatory factor analyses~\cite{thompson2004exploratory} to ensure the modified versions of our scales for reliability, technical competence, understandability, and personal attachment continued to exhibit the high degrees of reliability previously reported. We also conducted exploratory factor analysis~\cite{henson2006use} on our scale of trustworthiness to ensure each item measures the same underlying construct. All scales exhibited a high degree of reliability ($\alpha > 0.70$) except for understandability; thus, we exclude this scale from our analysis~\footnote{We did not find evidence of a significant lack of understanding of the scenarios. Ratings of understandability fell in the middle of the scale (M (SD) = 2.4 (.72) of 4), similar to how other trust facets were rated.}.

Next, we computed outcome scores for each scenario for trustworthiness, reliability, technical competence, and personal attachment. We then used an ANOVA model to understand the relationships between the scenario factors and these trust outcomes.

Finally, we performed a qualitative analysis to understand responses to our open-ended question about trustworthiness. We used an open coding approach~\cite{khandkar2009open} to determine high-level themes, followed by clustering responses into themes to determine overall trends. By mixing qualitative and quantitative methods, we aim to provide a more holistic picture of how each scenario factor affected trust in an AI-infused decision-making process.

\subsection{Stakes}
Participants' perceptions of the stakes of each scenario generally matched our intention. Participants rated the meal planning scenario as lower stakes (M (SD) = 3.7 (1.7) of 7) than the prison scenario (M (SD) = 6.3 (1.3) of 7), $F [1,630] = 438.2$, $p < .001$.

\subsection{Impact of Scenario Factors on Trust}
Factor analysis~\cite{thompson2004exploratory,henson2006use} indicates that our trustworthiness scale had a high level of reliability~\footnote{Our original trustworthiness scale contained 6 items, but exploratory factor analysis indicated that two of the items were not reliable. Thus, our final scale contains only the 4 items listed in Table~\ref{tab:scales}.}. In addition, confirmatory factor analysis indicated that although we had dropped and/or reworded items from the existing scales of reliability, technical competence, and personal attachment~\cite{madsen2000measuring, hoffman2018metrics}, these scales remained highly reliable.

We constructed an ANOVA model to test for main effects and all two-way interactions of the scenario factors to understand their effect on trust. As participants provided ratings of two scenarios, we included participant ID in the model as a random effect to control for this repeated measure. When presenting ANOVA results, we report effect sizes using partial $\eta^2$, which is the proportion of variance accounted for by each of the main effects or interactions in our model, controlling for all other effects~\footnote{Miles \& Shevlin~\cite{miles2001applying} advise that a partial $\eta^2$ of $\geq .01$ corresponds to a small effect, $\geq .06$ to a medium effect, and $\geq .14$ to a large effect.}. We present ANOVA results for all main effects in Table~\ref{tab:trust_facets}. However, for each of the trust facets (excluding reliability), two interaction terms were consistently significant: decider $\times$ trainer and decider $\times$ confidence. Therefore, we focus our analysis first on the non-interacting factors (stakes, interpretability, train/test description, and social transparency) and then around the interacting ones (decider, trainer, and confidence).

\subsubsection{Overall Ratings of Trust}
Ratings of the four trust facets fell in the middle of the scales (trustworthiness M (SD): 2.2 (.73), reliability: 2.6 (.73), technical competence: 2.4 (.83), personal attachment: 2.2 (.89) of 4), indicating possible skepticism of the use of AI in the scenarios. A few participants expressed strong opinions either for or against the use of AI in the scenarios.

\begin{quote}
    \textit{``I know human judgement can make [a] mistake, but I'm willing to take that chance. I don't want to take that chance with AI even if the result for me might be positive.'' (P127, stakes=high)}
\end{quote}

\begin{quote}
    \textit{``It is trustworthy because AI is fair, so it will only make [a] decision based on facts without being swayed.'' (P42, stakes=high)}
\end{quote}

\begin{quote}
    \textit{``I don't really like the idea of trusting a robot to make moral decisions for a human being.'' (P218, stakes=high)}
\end{quote}

\begin{quote}
    \textit{``I would completely trust the AI to make the right choices for me. I would feel confident the AI would do a good job.'' (P131, stakes=low)}
\end{quote}

\begin{table*}[ht]
    \small
    \centering
    \begin{tabular}{p{2cm}lllll|lllll} 
        \textbf{Factor} & \multicolumn{5}{c}{\textbf{Trustworthiness}} & \multicolumn{5}{c}{\textbf{Reliability}} \\
        \hline
            & M (SD) / M (SD) & $df$ & $F$ & $p$ & Partial $\eta^2$ 
            & M (SD) / M (SD) & $df$ & $F$ & $p$ & Partial $\eta^2$ \\
        \hline
        Stakes & & & & & & & & \\ 
        \hspace{0.2cm}\emph{low / high}.       & 2.4 (.69) / 2.0 (.72) & 1,303.5 & 51.4 & \textbf{< .001} & \textbf{.08} 
                                               & 2.6 (.70) / 2.6 (.76) & 1,289.3 &  .74 & n.s.            & < .01 \\
        Decider & & & & & & & & \\ 
        \hspace{0.2cm}\emph{human / ai}.       & 2.3 (.72) / 2.2 (.73) & 1,595.2 &  2.2 & n.s.   & < .01
                                               & 2.5 (.71) / 2.7 (.75) & 1,610.8 &  4.4 & .04    & < .01 \\
        Trainer & & & & & & & & \\ 
        \hspace{0.2cm}\emph{human / autoai}    & 2.3 (.73) / 2.1 (.71) & 1,605.1 &  9.8 & .002   &   .02
                                               & 2.6 (.72) / 2.6 (.75) & 1,561.7 &  2.8 & .09    & < .01 \\
        Interpretability & & & & & & & & \\ 
        \hspace{0.2cm}\emph{interp. / blckbx.} & 2.4 (.72) / 2.1 (.69) & 1,606.1 & 47.1 & \textbf{< .001} & \textbf{.07}
                                               & 2.8 (.69) / 2.4 (.74) & 1,577.3 & 31.8 & \textbf{< .001} & \textbf{.06} \\
        Train/Test Desc. & & & & & & & & \\ 
        \hspace{0.2cm}\emph{absent / present}  & 2.0 (.70) / 2.4 (.70) & 1,605.8 & 66.4 & \textbf{< .001} & \textbf{.10} 
                                               & 2.5 (.75) / 2.7 (.70) & 1,572.7 & 20.9 & < .001          & .04 \\
        Transparency & & & & & & & & \\ 
        \hspace{0.2cm}\emph{absent / present}  & 2.2 (.71) / 2.3 (.74) & 1,595.9 &  5.3 & .02  & .01 
                                               & 2.6 (.72) / 2.7 (.75) & 1,610.9 &  4.6 & .03  & .01 \\
        Confidence & & & & & & & & \\ 
        \hspace{0.2cm}\emph{absent / present}  & 2.2 (.71) / 2.3 (.74) & 1,605.4 &  2.9 & .09  & < .01 
                                               & 2.6 (.75) / 2.6 (.72) & 1,566.4 &  2.3 & n.s. & < .01 \\
        \hline
        \\
         & \multicolumn{5}{c}{\textbf{Technical Competence}} & \multicolumn{5}{c}{\textbf{Personal Attachment}} \\
        \hline
            & M (SD) / M (SD)  & $df$ & $F$ & $p$ & Partial $\eta^2$ 
            & M (SD) / M (SD)  & $df$ & $F$ & $p$ & Partial $\eta^2$ \\
        \hline
        Stakes & & & & & & & & \\ 
        \hspace{0.2cm}\emph{low / high}.       & 2.7 (.77) / 2.2 (.81) & 1,300.0 & 90.1 & \textbf{< .001} & \textbf{.15} 
                                               & 2.4 (.85) / 1.9 (.86) & 1,300.4 & 71.0 & \textbf{< .001} & \textbf{.12} \\
        Decider & & & & & & & & \\ 
        \hspace{0.2cm}\emph{human / ai}.       & 2.5 (.81) / 2.4 (.85) & 1,608.3 &  2.6 & .10  & < .01
                                               & 2.3 (.88) / 2.1 (.89) & 1,607.6 &  6.3 & .01  & .01 \\
        Trainer & & & & & & & & \\ 
        \hspace{0.2cm}\emph{human / autoai}    & 2.6 (.81) / 2.3 (.83) & 1,581.2 & 27.3 & < .001 & .05
                                               & 2.3 (.87) / 2.1 (.90) & 1,578.8 & 12.5 & < .001 & .02 \\
        Interpretability & & & & & & & & \\ 
        \hspace{0.2cm}\emph{interp. / blckbx.} & 2.6 (.81) / 2.2 (.81) & 1,590.3 & 38.5 & \textbf{< .001} & \textbf{.07}
                                               & 2.4 (.87) / 2.0 (.85) & 1,588.8 & 54.6 & \textbf{< .001} & \textbf{.10} \\
        Train/Test Desc. & & & & & & & & \\ 
        \hspace{0.2cm}\emph{absent / present}  & 2.2 (.80) / 2.6 (.80) & 1,587.6 & 61.7 & \textbf{< .001} & \textbf{.11}
                                               & 1.9 (.82) / 2.4 (.89) & 1,586.9 & 62.5 & \textbf{< .001} & \textbf{.11} \\
        Transparency & & & & & & & & \\ 
        \hspace{0.2cm}\emph{absent / present}  & 2.3 (.80) / 2.5 (.85) & 1,609.6 & 10.8 &   .001 & .02
                                               & 2.1 (.85) / 2.3 (.91) & 1,608.9 &  6.3 &   .01  & .01 \\
        Confidence & & & & & & & & \\ 
        \hspace{0.2cm}\emph{absent / present}  & 2.4 (.81) / 2.5 (.85) & 1,584.0 &  2.5 & n.s.   & < .01
                                               & 2.1 (.87) / 2.3 (.90) & 1,583.0 &  6.5 &.  .01  & .01 \\
        \hline
    \end{tabular}
    \caption{Effect of scenario factors on trustworthiness. Only main effects are shown, although significant interactions were present between decider $\times$ trainer and decider $\times$ confidence for each of the trust facets. Due to the inclusion of a random effect, degrees of freedom are fractional and vary for each factor. Bold text indicates significant, medium-to-large effects ($p \leq .05$ and partial $\eta^2 \geq .06$).}
    \label{tab:trust_facets}
\end{table*}

\subsubsection{Stakes}
Stakes exhibited significant and large effects on trustworthiness, technical competence, and personal attachment. The use of AI in lower-stakes scenarios vs. higher-stakes scenarios was associated with higher trustworthiness, higher technical competence, and higher personal attachment, shown in Table~\ref{tab:trust_facets}. In their own words, many participants expressed having less trust of AI in higher-stakes scenarios.

\begin{quote}
    \textit{``I simply don't trust any system that has input into important decisions. In this case, although the judge has the final say, he is considering input from a system that not only may be flawed, but may malfunction technically... I would hate the thought of this being applied to me and I would hope that the judge would use [her] wisdom and that any input from the AI would be minimal.'' (P93, stakes=high)}
\end{quote}

\begin{quote}
    \textit{``Since it's a less serious application like food, you don't have to worry so much about what data went into it or if it has weird biases or something like that.'' (P26, stakes=low)}
\end{quote}

Ratings of reliability did not differ between lower- and higher-stakes scenarios, indicating that participants may have felt the AI system described would produce consistent results, irregardless of the stakes of the decision being made.

\subsubsection{Model Interpretability}
Interpretability exhibited significant and strong effects on each of the trust facets. Interpretable models were consistently preferred over black box models: they were rated as more trustworthy, more reliable, and more technically appropriate than black box models, and participants liked decision processes more when they used interpretable models. Many participants expressed concerns over the use of black box models.

\begin{quote}
    \textit{``I would be scared if the process is applied on me... because one can not understand why it made this decision/recommendation.'' (P3, model=blackbox)}
\end{quote}

\begin{quote}
    \textit{``Most people including me would only trust an AI made and trained by professionals who are open to reveal the ways it works and its thought process like [how] normal humans would be able to say.'' (P15, model=blackbox)}
\end{quote}

\begin{quote}
    \textit{``Without that [interpretability] information, the AI could just be picking random items for all anyone knows.'' (P238, model=blackbox)}
\end{quote}

\begin{quote}
    \textit{``I don't trust this AI model because I cannot understand why and what factors caused it to make the recommendations.'' (P190, model=blackbox)}
\end{quote}

Although the statistical interaction was not significant, one participant did comment on how the use of interpretable models may be more important for higher-stakes decisions.

\begin{quote}
    \textit{``If the process was applied to myself I would probably see what food options the AI gave me before deciding how I felt about it. Being assigned a taco when you want pizza and not understanding why isn't nearly as bad not knowing why you were just sentenced to life in prison.'' (P66, model=blackbox)}
\end{quote}




\subsubsection{Train \& Test Set Description}
The presence of information about how a model was trained and tested had a significant and strong effect on ratings of trustworthiness, technical competence, and personal attachment. Decision-making processes that included this information were rated significantly higher than processes that did not include this information. The desire for transparency in how AI models are trained was discussed by many participants.

\begin{quote}
    \textit{``It is disconcerting that the data used to train the AI is not available.'' (P38, ttd=absent)}
\end{quote}

\begin{quote}
    \textit{``I would be more transparent with [the person subject to the decision] in how the system was developed and share test data with them to help them understand how the system works in order to instill confidence in them.'' (P4, ttd=absent)}
\end{quote} 

\begin{quote}
    \textit{``I do not like that there is no information available about how the model was trained. It makes me think they are hiding something.'' (P108, ttd=absent)}
\end{quote} 

\subsubsection{Social Transparency}
The availability of information about decisions made for others had significant, but small effects on all four trust facets. Decision-making processes that included this information were rated higher than processes that did not include this information. This desire to know what outcomes an AI-infused decision-making process produces for other people was clearly expressed by P40 and P199.

\begin{quote}
    \textit{``[This process] is untrustworthy because I do not get to see recommendations for other defendants.'' (P40, transparency=absent)}
\end{quote}

\begin{quote}
    \textit{``Lacking information about other recommendations that were made makes it so the judge can't see if the current recommendation is consistent with others.'' (P199, transparency=absent)}
\end{quote}

Although the quantitative analysis didn't show a significant interaction between stakes and social transparency, some participants expressed fewer concerns about not having social transparency in lower-stakes decisions.

\begin{quote}
    \textit{``Too little is know[n] about the decisions that this AI has already made for others. I wouldn't be too stressed out about it though, as it is only a meal that could be somewhat incorrect.'' (P145, transparency=absent, stakes=low)}
\end{quote}

\subsubsection{Decider}
Participants expressed mixed opinions toward who makes the ultimate decision in our scenarios: a human or an AI. Some felt comfortable with an AI making a decision, as long as that process was transparent.


\begin{quote}
    \textit{``This process is trustworthy because the most important parts of it are transparent. The information about how it makes its decision, how it is trained, and how much data it used to come to the decision is all available... I would feel comfortable with this process if it was applied to myself.'' (P116, decider=AI)}
\end{quote}

Other participants expressed preferences for having an AI make decisions due to it being free from human biases.

\begin{quote}
    \textit{``If the final decision is still made by the judge then it will be filled with the judges personal biases.'' (P6, decider=human)}
\end{quote}

\begin{quote}
    \textit{``It is trustworthy in the sense that someone being sentenced might get a reliably fair outcome vs a judge who might choose severe or lenient sentencing based on their feelings.'' (P64, decider=AI)}
\end{quote}

Conversely, some participants expressed concerns over trusting AI to make decisions because of their lack of empathy.

\begin{quote}
    \textit{``I feel like it isn't trustworthy because an AI lacks empathy and intuition to judge the likelihood of the perpetrator becoming a repeat offender.'' (P119, decider=AI)}
\end{quote}

\begin{quote}
    \textit{``I feel like there needs to be a human element, there's always additional circumstances that cold, hard numbers can't factor into when it comes to crime.'' (P33, decider=AI)}
\end{quote}

\begin{quote}
    \textit{``If this process were applied to me, then I would feel comfortable knowing that I would get a fair prison sentence based on the AI. However, I would also feel a bit uncomfortable, because there would be no way for me to obtain more leniency based on appeal to the judge.'' (P160, decider=human)}
\end{quote}

\subsubsection{Trainer}
Participants generally expressed a preference for AI models trained by data scientists rather than models trained using automated methods.

\begin{quote}
    \textit{``I think the way that the model is made would be slightly untrustworthy because the system trained itself.'' (P220, trainer=autoai)}
\end{quote}

\begin{quote}
    \textit{``It is not trustworthy to me because it was built completely without human intervention. I would trust it more if a person built it.'' (P263, trainer=autoai)}
\end{quote}

\begin{quote}
    \textit{``It is trustworthy. It's based on actual science data. It is created by scientists. I wouldn't change it at all.'' (P71, trainer=human)}
\end{quote}

\begin{quote}
    \textit{``If the AI wasn't built by a human, then I don't trust it.'' (P158, trainer=autoai)}
\end{quote}


There were significant, but very small interactions between decider and trainer for trustworthiness ($F [1,605.1] = 4.0$, $p = .05$, partial $\eta^2 < .01$), technical competence ($F [1,581.3] = 4.3$, $p = .04$, partial $\eta^2 < .01$), and personal attachment ($F [1,578.7] = 3.9$, $p = .05$, partial $\eta^2 < .01$). When the decider was an AI, all three facets were rated higher ($+0.2$ to $+0.4$ points) when humans were responsible for training the AI than when automated technologies were used; thus, participants favored human-trained AI when that AI is solely responsible for making a decision. When the decider was a human, there were no differences in trustworthiness or personal attachment based on how the AI was trained, but participants did feel that human-trained AI was more technically competent ($+0.15$ points) in this case.

\subsubsection{Model Confidence}
As with trainer, there were significant, but very small interactions between decider and confidence for trustworthiness ($F [1,596.9] = 7.3$, $p < .01$, partial $\eta^2 < .01$), technical competence ($F [1,573.8] = 5.0$, $p = .03$, partial $\eta^2 < .01$), and personal attachment ($F [1,573.3] = 4.7$, $p = .03$, partial $\eta^2 < .01$). When the decider was an AI, all three facets were rated higher ($+0.2$ to $+0.3$ points) when information about a model's confidence was present. When the decider was a human, differences in confidence scores were negligible. Thus, from a statistical perspective, information about a model's confidence seems highly important when an AI is responsible for making a decision, but less so when there is a human in the loop. However, in their own words, participants were strongly in favor of having information about a model's confidence.

\begin{quote}
    \textit{``[The use of AI is] completely inappropriate. Especially since the degree of confidence is left secret. This is a grotesquely unfair process.'' (P82, decider=AI, confidence=absent)}
\end{quote}

\begin{quote}
    \textit{``This is untrustworthy since there is no confidence, no information on the data, and no input on other decisions. I would change everything about this process except for having the judge involved.'' (P210, decider=human, confidence=absent)}
\end{quote}

\begin{quote}
    \textit{``It would be more trustworthy if it showed me its confidence rating... Of course with the final decisions being made by the chef, I expect the results to be good (assuming the chef is halfway competent).'' (P39, decider=human, confidence=absent)}
\end{quote}

In contrast, one lone participant expressed that a lack of confidence information was ``not a big deal'' in lower-stakes scenarios.

\begin{quote}
    \textit{``...info about the AI's confidence level of its recommendations not being readily available is not a big deal to me so it is acceptable... I would feel confident using it.'' (P167, stakes=low, confidence=absent)}
\end{quote}



\section{Discussion}
Trust is a complex idea, irreducible to just one ``thing.'' There are many reasons why we trust something: when we expect to rely on it~\cite{rotter1967new}, when we believe it helps us achieve our goals~\cite{lee2004trust}, when we heed its advice.~\cite{madsen2000measuring}, and as stated by many of our participants, when it is transparent in its operation and we understand how it works.

We have examined several aspects of trust across a wide range of AI-infused decision-making scenarios in order to ascertain what factors are most important for establishing trust. We see the largest statistical effects come from interpretability -- interpretable models were favored over black box models -- and the inclusion of information about how an AI model was trained and tested. But, our quantitative analysis hides the more nuanced, ``messier'' side of trust, in which peoples' opinions are complex and sometimes conflicting.

We discuss a number of additional themes that emerged from our analysis of the qualitative feedback, such as the role of human empathy in higher-stakes scenarios and AI's inability to factor in mitigating circumstances on which it may not have been trained.



\subsection{`Black Boxes' are Less Trustworthy}
In line with our intuition, we found that the use of AI was more accepted when making lower-stakes decisions than higher-stakes ones. In both cases, the most important factor for establishing trust was the use of interpretable models. Participants simply did not trust the use of black box models as much as interpretable models, clearly expressed by P90.

\begin{quote}
    \textit{``A black box AI that determines prison sentences? There is no way anyone should allow this without it being open-source and publicly scrutinized.'' (P90, model=blackbox)}
\end{quote}

This result agrees with recent arguments by Rudin~\cite{rudin2018please}, who expresses strong opinions in favor of using interpretable models over black box models for high-stakes decisions. Based on our findings, we extend this guidance to cover lower-stakes decisions as well.

\subsection{In Transparent Models We Trust}
The inclusion of information about how an AI model was trained and tested had a significant and large effect on trust. Participants clearly preferred having transparency in how a model was created, in both lower- and higher-stakes scenarios, and they were especially passionate that such information is included for higher-stakes decisions.

\begin{quote}
    \textit{``If it doesn't share its confidence or how it comes to its sentencing, where the data came from that it was trained with, then how will people be able to trust it?'' (P75, ttd=absent)}
\end{quote}

This result echoes recent calls for including a set of factual information about how AI models were trained and tested when distributing those models~\cite{hind2018increasing, mitchell2019model}. Our results provide statistical evidence that trust can be improved in this manner.

\subsection{Who Gets to Decide?}
We found interesting, conflicting opinions about the entity making the ultimate decision in our scenarios. Some participants were comfortable with AI making a decision, so long as it provided its level of confidence. Other participants felt that the use of AI was ``completely inappropriate... especially [when] the degree of confidence is left secret'' (P82).

Who trained the model also seems to matter: when AI makes the decision, participants had more trust when that AI was trained by people, but when a human makes the decision, there was no difference in trust when that AI was trained by people or using automated methods. Thus, the deeper conclusion seems to be that a human-in-the-loop is important \emph{somewhere} in the process -- either during training time or at decision time -- in order to increase trust.

\subsection{Trust of Automated AI}
Within data science, there is an increasing trend toward automation, especially of the labor-intensive phases of data cleaning, feature engineering, and model building~\cite{wang2019humanai}. Even though the performance of models created with automated methods can be on par (or exceed) that of models created by human data scientists~\cite{yao2018autoai}, participants tended to be skeptical of the use of automated methods.

\begin{quote}
    \textit{``I think the way that the model is made would be slightly untrustworthy because the system trained itself.'' (P220, trainer=autoai)}
\end{quote}

In addition to skepticism of automated methods, some participants were skeptical of human data scientists as well.

\begin{quote}
    \textit{``For all we know the scientists could train the AI to do something that's entirely unfair, but there's no way to judge that because there's zero access to information.'' (P309, trainer=human)}
\end{quote}

\begin{quote}
    \textit{``[The data scientists] may be very well-meaning but they can make errors because they are human.'' (P277, trainer=human)}
\end{quote}

In contrast, other participants felt that human data scientists could implicitly be trusted.

\begin{quote}
    \textit{``I trust the data scientists who created the algorithm to know what they're doing.'' (P257, trainer=human)}
\end{quote}

\begin{quote}
    \textit{``I think its trustworthy because the programmers programmed it that way.'' (P345, trainer=human)}
\end{quote}


\subsection{Mitigating Human Biases}
Historical data often carries biases, of which we may not even be aware, because of the existing human and systemic biases present in the processes that produced that data. The dangers of using historical data to train AI was succinctly captured by P100.

\begin{quote}
    \textit{``When you vaguely base decision making only on historical data, what you're bound to get is history repeating itself.'' (P100)}
\end{quote}

Many participants expressed such concerns over training on biased data, especially for higher-stakes decisions.


\begin{quote}
    \textit{``The AI uses historical data. In this country, we know that that data does not necessarily represent fair practices. Biases ha[ve] always occurred against a segment of the population and this AI would just perpetuate those biases.'' (P213, stakes=high)}    
\end{quote}

\begin{quote}
    \textit{``The potential for the machine to become biased is extremely high, considering there's sufficient evidence to suggest that decades of criminal convictions have been heavily biased in the court system throughout history.'' (P104, stakes=high)}
\end{quote}

These sentiments reflect the need for specific steps to be taken when training AI systems to mitigate and minimize the impact of such biases. Bellamy et al.~\cite{bellamy2019aif} provide additional discussion on this topic.

Despite these concerns around the use of biased historical data to train an AI, AI was simultaneously seen as an impartial and fair decider compared to people.

\begin{quote}
    \textit{``I would feel good because it would be more impartial than a human.'' (P163, decider=AI)}
\end{quote}

\begin{quote}
    \textit{``If the final decision is still made by the judge then it will be filled with the judges personal biases.'' (P6, decider=human)}
\end{quote}

\subsection{AI's Lack of Empathy and Morality}
Participants expressed conflicting concerns about the lack of empathy in an AI decider. Some participants felt that this was a positive aspect that increased trust in the process by keeping human emotions in check.

\begin{quote}
    \textit{``Having input from AI will lessen other emotional factors that might affect the judge.'' (P40, stakes=high, decider=human)}
\end{quote}

\begin{quote}
    \textit{``Someone being sentenced might get a reliably fair outcome vs a judge who might choose severe or lenient sentencing based on their feelings.'' (P64, stakes=high, decider=AI)}
\end{quote}

\begin{quote}
    \textit{``[The AI] eliminates human emotion and hands out conventions based on facts and unemotional things. Nothing should be changed. I would feel like I received a fair sentencing.'' (P251, stakes=high, decider=AI)}
\end{quote}

Other participants were wary of AI's lack of empathy and morality and felt that AI should not be used for making higher-stakes decisions.

\begin{quote}
    \textit{``AI is not capable of empathy and that needs to be a factor in this type of a decision. I would remove the AI component to make it a trustworthy process. I would feel scared and stressed if this process were applied to myself.'' (P19, stakes=high, decider=human)}
\end{quote}

\begin{quote}
    \textit{``I don't think that an AI, no matter how good the data input into it, can make a moral decision.'' (P299, stakes=high, decider=AI)}
\end{quote}

\begin{quote}
    \textit{``If this process were applied to me I would feel like my life is being gambled at a 50/50 chance by a thing that doesnt even have a moral compass nor compassion.'' (P306, stakes=high, decider=AI)}
\end{quote}

No participants discussed issues of empathy or morality with regard to the lower-stakes scenario. As summed up by P57,

\begin{quote}
    \textit{``It seems trustworthy as the worst that could happen is receiving a meal that you wouldn't enjoy.'' (P57, stakes=low)}
\end{quote}

\subsection{Inability to Consider Special Circumstances}
Participants expressed concerns over AI models not being able to factor in special considerations, mitigating factors, or circumstantial evidence that are not reflected in historical data.

\begin{quote}
    \textit{``A machine cannot process all of the information that it hasn't encountered before.'' (P98, stakes=high)}
\end{quote}

\begin{quote}
    \textit{``There may be other relevant factors to the case that the AI would not consider.'' (P10, stakes=high)}
\end{quote}

Some participants felt worried about ``generaliz[ing] across... different cases'' (P67)  and applying the process to ``a hypothetical version of myself'' (P68) due to the AI's inability to consider special circumstances for which it hasn't been trained.

\begin{quote}
    \textit{``The individual's specific circumstances, such as physical health and specific individual nutritional needs and energetics are not being taken into account.'' (P85, stakes=low, decider=AI)}
\end{quote}

\begin{quote}
    \textit{``There's always additional circumstances that cold, hard numbers can't factor into when it comes to crime.'' (P33, stakes=high, decider=AI)}
\end{quote}

\subsection{Designing Trustworthy AI}
Our qualitative results highlight the myriad of individual differences that exist around peoples' perceptions of the use of AI in decision making. In sum, there is no ``magic formula'' that can be prescribed to guarantee that an AI system is trusted. Rather, although trust is affected in some part by the factors we examined -- especially the use of interpretable models and the presence of information about how an AI model was trained and tested -- many more factors influence peoples' feelings, and sometimes in contradictory ways. Some participants felt skeptical of automated methods being used to train AI models, others questioned whether data scientists always had fair intentions. Some participants felt that AI's lack of empathy was a boon to its impartiality, others felt that lack of empathy to be limiting in its ability to handle special circumstances.

In light of these conflicting viewpoints, one message seems clear: designing trustworthy AI is a difficult task, and navigating it successfully will require deep and thoughtful input from not only the people building the AI, but from the people using it and from the people affected by it as well.

\section{Limitations and Future Directions}
We acknowledge several important limitations to our work that limit the generalizability and comprehensiveness of our findings. Chiefly, our sample consisted of native English speakers drawn from English-speaking countries on Mechanical Turk. Thus, we do not make any claims about the cross-cultural validity of our work. We also did not capture any information about our participants' background or familiarity with AI, and thus, opinions expressed may be based on faulty assumptions as to what AI can (or cannot) do.

Our selection of scenarios was designed to capture opinions on two ends of a spectrum of consequences: a lower-stakes scenario in which making a wrong decision may be inconvenient, and a higher-stakes scenario in which making a wrong decision may be catastrophic. Although we do see important differences in trust between these two scenarios, we acknowledge that they may not represent the full spectrum of how AI is being incorporated into decision-making processes. We urge AI system designers to carefully consider their needs for trust, and the mechanisms by which that trust is established, when designing AI-infused decision-making processes.

In addition, although we attempted to identify a wide range of important factors that may impact trust of an AI-infused decision-making process, there may exist other factors that we did not consider. For example, we only made comparison between interpretable and black box models. Numerous algorithms are being developed to \emph{explain} black box models in a way that makes them more akin to interpretable models (see~\cite{arya2018aix} for a comprehensive review; c.f.~\cite{rudin2018please} for limitations of these approaches). However, these explanations are typically developed with the data scientist or AI engineer in mind, and may not be appropriate for consumption by the person who is ultimately affected by the AI's output. Additional work is needed to understand the abilities of different types of explanations in building trust in an AI system from a multitude of perspectives -- the people building the system, the people using the system, and the people affected by it.

Finally, our examination of the presence or absence of information about how an AI model was trained and tested was similarly high-level. Additional work is needed to understand \emph{what} information is actually required to establish trust amongst the different stakeholders in an AI system's lifecycle. For example, data scientists may need a different set of facts to establish trust in an AI model than the sponsor or executive in charge of deciding whether that model is deployed. Governments may have their own requirements about what information must be disclosed about an AI model (see~\cite{hleg2019ethics}). Additional work is needed to uncover these information requirements and understand their impact on trust.

\section{Conclusion}
With the growing usage of AI and machine learning models in making decisions that significantly impact our lives, the need to understand how to build trusted AI is paramount. Although the need for trust in AI systems has always been present, new applications of AI for making higher-stakes decisions, such as determining whether we are approved for a home loan or admitted to college, exacerbate issues of trust.

Our study is an attempt to provide clear, quantified guidance for how to increase trust in AI-infused decision-making processes. We find that certain features, such as the use of interpretable models and the inclusion of information about how an AI model was trained and tested, are associated with significant improvements to peoples' ratings of trust. However, these results do not tell the whole story. For some people, the mere inclusion of AI in a decision-making process was completely inappropriate; for others, they felt that AI provided a degree of impartiality that guarded against human biases.

It seems inevitable that AI systems will become a larger part of the decisions we make, and the decisions made for us, in our lives. Although our interpretation of the results of this study has been positive (i.e., more trust is always desirable), we urge researchers and AI system designers to also consider the converse: when ought people be mistrustful of AI, and how can we design mechanisms for people to \emph{distrust} AI in those situations?





\bibliographystyle{ACM-Reference-Format}
\bibliography{bibliography}

\end{document}